\documentclass[11pt]{article}
\pdfoutput=1
\usepackage[utf8]{inputenc}
\usepackage[left=20mm,right=20mm,top=25mm,bottom=25mm]{geometry}
\usepackage[T1]{fontenc}

\usepackage[affil-it]{authblk}
\usepackage{amssymb}

\usepackage{lineno}

\usepackage[ngerman, main=english]{babel}
\usepackage[babel]{csquotes}

\usepackage{listings} 

\usepackage{booktabs} 
\newcommand{\ra}[1]{\renewcommand{\arraystretch}{#1}} 
\usepackage{float} 
\usepackage{multirow}
\usepackage{subcaption}

\usepackage[hidelinks]{hyperref}
\def\equationautorefname~#1\null{Eq. (#1)\null}

\makeatletter
\newcommand*{\rom}[1]{\expandafter\@slowromancap\romannumeral #1@} 
\makeatother

\usepackage{hologo}

\usepackage{pgfplots}
\pgfplotsset{compat=newest} 
\pgfplotsset{plot coordinates/math parser=false} 

\pgfplotsset{compat=newest}
\pgfplotsset{plot coordinates/math parser=false}

\newlength\figureheight				
\newlength\figurewidth				
\newlength\figureheightbig			
\newlength\figurewidthbig			
\pgfplotsset{major grid style={%
				dashed,			
				black,			
				}}

\pgfplotsset{tick align = inside}	
\pgfplotsset{tick style={
				line width=0.2mm,	
				black				
				}}	

\pgfplotsset{every axis/.append style={
				line width = 0.2mm		
				}}

\pgfplotsset{every axis plot/.append style={line width=0.5pt}} 

\usetikzlibrary{shapes.geometric} 
\usepackage{smartdiagram}

\usepackage{mathtools} 

\definecolor{tud1a}{HTML}{5D85C3}
\definecolor{tud2atrue}{HTML}{009CDA}
\definecolor{tud2a}{rgb}{0.00000,0.44700,0.74100}%
\definecolor{tud3a}{HTML}{50B695}
\definecolor{tud3ai}{HTML}{009a67} 
\definecolor{tud4a}{HTML}{AFCC50}
\definecolor{tud5a}{HTML}{DDDF48}
\definecolor{tud6a}{HTML}{FFE05C}
\definecolor{tud6b}{HTML}{FDCA00}
\definecolor{tud7b}{HTML}{F5A300} 
\definecolor{tud8a}{HTML}{EE7A34}
\definecolor{tud8b}{HTML}{EC6500}
\definecolor{tud9a}{HTML}{E9503E}
\definecolor{tud11b}{HTML}{721085}
\definecolor{tud1b}{HTML}{005AA9} 
\definecolor{tud3b}{HTML}{009D81}
\definecolor{tud4b}{HTML}{99C000}
\definecolor{tud4c}{HTML}{7FAB16}
\definecolor{tud4d}{HTML}{6A8B22}
\definecolor{tud9b}{HTML}{E6001A}
\definecolor{tud9c}{HTML}{B90F22} 
\definecolor{tud10a}{HTML}{A60084} 
\definecolor{sourcecode}{HTML}{E9E9E9}
\definecolor{mygrey}{HTML}{C0C0C0}

\newcommand*\diff{\mathop{}\!\mathrm{d}}

\newcommand{\Arch}{\operatorname{\mathit{A\kern-.06em r}}} 
\newcommand{\Biot}{\operatorname{\mathit{B\kern-.06em i}}} 
\newcommand{\Cauc}{\operatorname{\mathit{C\kern-.07em a}}} 
\newcommand{\Damk}{\operatorname{\mathit{D\kern-.06em a}}} 
\newcommand{\Eule}{\operatorname{\mathit{E\kern-.03em u}}} 
\newcommand{\Four}{\operatorname{\mathit{F\kern-.10em o}}} 
\newcommand{\Frou}{\operatorname{\mathit{F\kern-.07em r}}} 
\newcommand{\Gras}{\operatorname{\mathit{G\kern-.05em r}}} 
\newcommand{\Karl}{\operatorname{\mathit{K\kern-.11em a}}} 
\newcommand{\Knud}{\operatorname{\mathit{K\kern-.11em n}}} 
\newcommand{\Lewi}{\operatorname{\mathit{L\kern-.05em e}}} 
\newcommand{\Mach}{\operatorname{\mathit{M\kern-.10em a}}} 
\newcommand{\Nuss}{\operatorname{\mathit{N\kern-.09em u}}} 
\newcommand{\Pecl}{\operatorname{\mathit{P\kern-.08em e}}} 
\newcommand{\Pran}{\operatorname{\mathit{P\kern-.03em r}}} 
\newcommand{\Rayl}{\operatorname{\mathit{R\kern-.04em a}}} 
\newcommand{\Reyn}{\operatorname{\mathit{R\kern-.04em e}}} 
\newcommand{\Rich}{\operatorname{\mathit{R\kern-.06em i}}} 
\newcommand{\Schm}{\operatorname{\mathit{S\kern-.07em c}}} 
\newcommand{\Sher}{\operatorname{\mathit{S\kern-.07em h}}} 
\newcommand{\Stro}{\operatorname{\mathit{S\kern-.07em r}}} 
\newcommand{\Webe}{\operatorname{\mathit{W\kern-.14em e}}} 

\usepgfplotslibrary{external} 
\tikzsetexternalprefix{TikzFig/}

\usepackage{cuted}

\input{tudacolors.def}
\usepackage[maxbibnames=10, maxcitenames=3, giveninits=true,
    natbib=true, bibencoding=utf8, isbn=false, url=false, backend=bibtex, style=numeric-comp]{biblatex} 
\AtEveryBibitem{\clearfield{month}}
\AtEveryBibitem{\clearfield{day}}
\AtEveryBibitem{\clearlist{language}}
 
 \usepackage{hyperref} 
\hypersetup{
    colorlinks=true,
    linkcolor=tud2a,
    filecolor=magenta,      
    urlcolor=tud2a,
    citecolor=tud2a,
    pdfpagemode=FullScreen,
    }
    
\begin{document}




\title{
Endeavouring Intelligent Process Self-Control by Employing Digital Twin Methodology: Proof-of-Concept Study for Cooking Applications
}

\author[1,*]{Maximilian~Kannapinn}
\author[1]{Michael~Schäfer}
\affil[1]{\footnotesize Institute for Numerical Methods in Mechanical Engineering \& Graduate School of Computational Engineering, \protect \\
Department of Mechanical Engineering \& Centre for Computational Engineering,\protect \\ Technical University of Darmstadt, Dolivostr.~15, 64293 Darmstadt, Germany}
\affil[*]{\footnotesize Corresponding author, Email: \href{mailto:research@maxkann.de}{research@maxkann.de}}

\date{September 10, 2019}

\maketitle

 \par\noindent\rule{\textwidth}{0.4pt}           
\begin{abstract} \noindent
This work demonstrates the use of the Digital Twin methodology to predict the water concentration and temperature of chicken meat. It marks a milestone on the path to autonomous cooking devices that do not only control device temperatures but culinary food quality markers as well.
A custom water transport equation is coupled to the energy equation. The transport equations are implemented in ANSYS Fluent 2019R2 via User Defined Function (UDF) code written in C. The model is in good agreement with experiments provided by project partners. Thermal fluid-structure interaction simulations of pan-frying are performed to obtain realistic heat transfer coefficients. They indicate that the coupling of food transport equations to the surrounding heat transfer mechanisms, such as radiation and natural convection, seems promising for future research.

Co-simulation of the process is not feasible during operation in the field, so reduced-order models (ROM) are introduced. An evaluation of ROM toolkits on the ANSYS platform reveals that linear time-invariant (LTI) models are unsuitable for cooking applications. In contrast, the recently launched Dynamic ROM Builder predicts the core temperatures with significantly low errors (factor ten below the model and discretization errors of the full-order model). 
Two examples demonstrate the usage of a Digital Twin controlling the core temperature of chicken fillets. The PI closed-loop control system remains insensitive to errors induced by the Dynamic ROM evaluation. 

\end{abstract}

\vspace*{0.5ex}
{\textbf{Key words:} Digital Twin, Thermal Fluid-Structure Interaction, Process Control, Reduced-Order Modelling
\par\noindent\rule{\textwidth}{0.4pt}\vspace*{2pt}
{\small
Accepted version of the manuscript published in \emph{CADFEM ANSYS Simulation Conference 2019}. \\
Received a best paper award.\\
Date accepted: September 10, 2019. 
License: \href{https://creativecommons.org/licenses/by-nc-nd/4.0/legalcode}{CC BY-NC-ND 4.0}
}
\vspace*{-1.6mm}
\par\noindent\rule{\textwidth}{0.4pt}

%
%



\section{The Digital Twin for cooking applications}

For decades product development has profited from simulated data that cannot easily or at all be obtained by measurements. Recent developments in Digital Twin methodology enable us to leverage this information during the operational phase. 
This proof-of-concept focuses on a cooking process, specifically the roasting and pan-frying of chicken meat. The presented method can be interpreted as a virtual chef of an intelligent cooking device. It can make decisions based on temperature measurements in the device to control the pre-defined final culinary quality of the food. The primary state variables of the cooking process of chicken meat are the mass-weighted water concentration and temperature. Two coupled transport equations are implemented in ANSYS Fluent and validated with experimental data.

\section{Modelling of a cooking process}

The general conservation equation for the water concentration $C$ reads
\begin{align}
\label{eq:conservation}
\frac{\diff}{\diff t} \int_\Omega C \diff \Omega = \int_\Gamma J \diff \Gamma + \int_\Omega S \diff \Omega
\end{align}
with fluxes $J$ over the domain boundaries $\Gamma$ and sources $S$ in the domain $\Omega$. We apply Reynold's transport theorem~\cite{cfd_schaefer} on the left side. This introduces the convective flux term, where $u_i$ denotes water velocity. For $J$ we follow the convention that inward-facing fluxes $j_k$ are positive over a boundary with outward-facing normal vectors $n_k$:
\begin{align}
\label{eq:conservation-rtt}
\int_\Omega \left[ \frac{\partial C}{\partial t} + \frac{\partial}{\partial x_i} \left( C\,u_i\right)\right] \diff \Omega = \int_\Gamma - j_k \, n_k \diff \Gamma + \int_\Omega S \diff \Omega\,.
\end{align}
Fick's law of diffusion
\begin{align}
j_i = -D_{ik}\,\frac{\partial C}{\partial x_k}
\end{align}
characterizes particle flow in the opposite direction of the steepest ascent of the field quantity with a diffusion coefficient ${D}_{ik}$
~\cite{thermo_baehr_wust2013}, which is a second-order tensor usually containing positive values in a diffusion scenario. After using Gauss's divergence theorem on the first term of the right-hand side of \autoref{eq:conservation-rtt}, we can switch to a differential representation of the transport equation:
\begin{align}
\label{eq:conservation-differential}
\frac{\partial C}{\partial t} = \frac{\partial}{\partial x_i} \left( D_{ik} \frac{\partial C}{\partial x_k} \right)- \frac{\partial}{\partial x_i} \left( Cu_i\right)+ S\,.
\end{align}
Darcy's law is used to relate the velocity $u_i$ to a pressure gradient $\nabla p$: 
\begin{align}\label{eq:ui}
u_i = -\frac{\kappa}{\mu_w}\,\frac{\partial p}{\partial x_i} .
\end{align}
$\mu_w$ denotes the dynamic viscosity of water. 
The permeability $\kappa$ of chicken meat is rarely found within the literature. It was experimentally obtained by Datta~\cite{cm_datta_permeability2006} to be $\kappa \in [1\mathrm{e}{-17},1\mathrm{e}{-19}]\,\text{m}^2$. 
Feyissa evaluated values of $\kappa \in [1\mathrm{e}{-16},1\mathrm{e}{-17}]\,\text{m}^2$ and suggests $\kappa =1\mathrm{e}{-17}\,\text{m}^2$ to match experimental data of raw meat. Too high values can lead to an over-prediction of water concentration rise in the center of the probe~\cite{cm_feyissa_3D2013}. We follow the parameter settings $\kappa=3\,\mathrm{e}{-17}\,\text{m}^2$ of Rabeler~\cite{cm_rabeler_mod2018} to enable comparison in Section~\ref{sec:case1}. 

Protein denaturation induces shrinkage during a roasting process, which results in water exudation at the surface.  
Meat science researchers related this phenomenon to an adaption of the Flory-Rehner theory~\cite{cm_floryrehner_polymerswelling1943}, which describes the swelling pressure in polymer gels~\cite{cm_quesadaperez_gelswelling2011}.
For cooking applications, $p$ can be related to the water concentration in meat~\cite{cm_datta_porousmedia06a,cm_vandersman_moisture2007a,cm_vandersman_chickentunnel2013,cm_feyissa_3D2013,cm_rabeler_mod2018} as
\begin{align}\label{eq:p}
p = G^\prime \left( C - C_{eq}(T)\right).
\end{align}
The induced pressure can be modeled to be proportional to the difference between the local water concentration $C$ and the equilibrium water concentration $C_{eq}(T)$, also known as water holding capacity.
$G^\prime(T)$ models a storage modulus of the food to be dependent on the temperature. Both quantities have been fitted as sigmoidal functions to match experimental data~\cite{cm_vandersman_moisture2007a,cm_rabeler_kin2018}.
The water holding capacity $C_\text{eq}$ reads:
\begin{align}
C_\text{eq} &= y_{w,0} - \frac{0.31}{1+ 30.0\, \text{exp}\left(-0.17 \left(T-T_\sigma\right)\right) }\,,\\
y_{w,0} &= 0.77\,,\\
T_\sigma &=315 \text{ K}\,.
\end{align}
The storage modulus $G^\prime(T)$ is modeled as:
\begin{align}
G^\prime &= G^\prime_\text{max} + \frac{G^\prime_0 - G^\prime_\text{max}}{1 + \text{exp} \left(\frac{T-\overline{T}}{\Delta T}\right) }\,,\\
G^\prime_\text{max} &= 92\, 000 \text{ Pa}\,,\\
G^\prime_\text{0} &= 13\, 500 \text{ Pa}\,,\\
\overline{T} &= 342.15 \text{ K}\,,\\
\Delta T &= 4 \text{ K}\,.
\end{align}
Inserting~\autoref{eq:p} in \autoref{eq:ui} and ignoring the contribution of $\frac{\partial G^\prime}{\partial x_i}$, we can relate $u_i$ to $C$ as 
\begin{align}
u_i = -\frac{\kappa \, G^\prime}{\mu_w}\,\frac{\partial}{\partial x_i} \left[ C - C_{eq}(T)\right].
\end{align}
Analogous to the mass balance of $C$, we can derive energy conservation and reformulate it to a transport equation:
\begin{align}\label{eq:energy}
c_p \, \rho \frac{\partial T}{\partial t} = \frac{\partial}{\partial x_i} \left(\lambda_{ik} \, \frac{\partial T}{\partial x_k}\right) - \rho \, c_p \,u_k \frac{\partial T}{\partial x_k}\,,
\end{align}
where $c_p$ denotes the specific heat capacity of chicken meat and $\rho$ is the density of water.

\subsection{Implementation via User Defined Functions}

The User Defined Scalar (UDS) capability of ANSYS Fluent enables the implementation of custom transport equations which will be solved in a segregated manner. Each term in the transport equation is coded separately with User Defined Functions (UDF). 
The unsteady term is split into a discrete finite difference~\cite{ansys_udf_2019R1} as
\begin{align}
-\int \frac{\partial}{\partial t} C \diff V \approx -\left[ \frac{C^n-C^{n-1}}{\Delta t}\right]\,\Delta V = \underbrace{-\frac{\Delta V}{\Delta t}}_{\text{apu}} C^n +\underbrace{ \frac{\Delta V\,C^{n-1}}{\Delta t}}_{\text{su}}
\end{align}
and is handed over to Fluent via Algorithm~\ref{c-unsteady}.
\lstdefinestyle{customc}{
 belowcaptionskip=1\baselineskip,
 breaklines=true,
 frame=single,
 xleftmargin=1\parindent,
 xrightmargin=1\parindent,
 language=C,
 showstringspaces=false,
 basicstyle=\footnotesize\ttfamily,
 keywordstyle=\bfseries\color{black},
 commentstyle=\itshape\color{TUDa-0c},
 identifierstyle=\color{black},
 stringstyle=\color{black},
}
\lstset{style=customc}
\begin{lstlisting}[language=C, firstline=9,label={c-unsteady},caption={UDF for unsteady term in $C$ transport equation.}]
DEFINE_UDS_UNSTEADY(c_unsteady_nomass,c,t,i,apu,su)
{
real physical_dt, vol, rho, phi_old;
physical_dt = RP_Get_Real("physical-time-step");
vol = C_VOLUME(c,t);
// rho = C_R_M1(c,t);                           /*density of previous time step*/
*apu = -vol / physical_dt;                      /*implicit part without rho*/
phi_old = C_STORAGE_R(c,t,SV_UDSI_M1(i));       /*previous uds value*/
*su = vol*phi_old/physical_dt;                  /*explicit part without rho*/
}


\end{lstlisting}
The term $C - C_{eq}(T)$ is written to a separately defined UDS with a \texttt{DEFINE\_ADJUST()} routine including respective cell loops \texttt{thread\_loop\_c}, compare Algorithm~\ref{c-ceq}. The UDS acts as variable storage, and it is not solved actively. Hence, the needed gradient calculation of the term can be realized by calling Fluent's proper gradient calls, e.g., \texttt{C\_UDSI\_G(c,t,i)}. 

\begin{lstlisting}[language=C, firstline=11,label={c-ceq},caption={Manipulation of UDS-1 cell and boundary face values.}]
DEFINE_ADJUST(adjust_ceq_face, domain)
{
Thread *t;
cell_t c;
face_t f;
domain = Get_Domain(domain_ID);

/* Fill cells of UDS 1 with the c_old - c_eq. */
thread_loop_c (t,domain)
{
    begin_c_loop (c,t)
    {
        C_UDSI(c,t,1) =C_UDSI(c,t,0) - 0.77 + 0.31/(1.0+30.0*exp(-0.17*(C_UDSI(c,t,2)-315.0)));
    }
    end_c_loop (c,t)
}
/* Fill boundary faces of UDS 1 with the c_old - c_eq. */
thread_loop_f (t,domain)
{
if(BOUNDARY_FACE_THREAD_P(t))
{
if (THREAD_STORAGE(t,SV_UDS_I(1))!=NULL) // only boundaries where uds 1 exists
{
begin_f_loop (f,t)
    {
    F_UDSI(f,t,1) = F_UDSI(f,t,0) - 0.77 + 0.31/(1.0+30.0*exp(-0.17*(F_UDSI(f,t,2)-315.0)));
    }
end_f_loop (f,t)
}}}}
\end{lstlisting}
Special attention has to be paid to domain surfaces. Supplementary boundary face loops \texttt{thread\_loop\_f} in the aforementioned \texttt{DEFINE\_ADJUST()} routine are necessary to ensure proper gradient calculation of $C - C_{eq}(T)$. 

The convective flux term is computed with Algorithm~\ref{c-flux}. On boundary faces, it is directly evaluated as all necessary variables are allocated here. For faces inside the domain, we average adjacent cell values for simplicity. The expected numerical error will be insignificant on the orthogonal meshes used in this study. For unstructured meshes, correction factors would have to be calculated for deformed cells~\cite{cfd_schaefer}.
\begin{lstlisting}[language=C, firstline=10,label={c-flux},caption={$C - C_{eq}(T)$ gradient evaluation for convective flux call.}]
DEFINE_UDS_FLUX(c_flux_uw,f,t,i)
{
cell_t c0, c1 = -1;
Thread *t0, *t1 = NULL;
real NV_VEC(grad_c), NV_VEC(A), flux = 0.0, gprime;
c0 = F_C0(f,t);
t0 = F_C0_THREAD(f,t);
F_AREA(A, f, t);

if (BOUNDARY_FACE_THREAD_P(t)) /* if face lies at domain boundary, use face values */
{
    /* get gradient of c-c_eq */
    NV_D(grad_c, =, C_UDSI_G(c0,t0,1)[0],C_UDSI_G(c0,t0,1)[1],C_UDSI_G(c0,t0,1)[2]);
    gprime=(gmax + (gzero-gmax)/(1.0+exp(C_UDSI(c0,t0,2)/4.0-342.15/4.0)));
    flux = -3E-17/0.988E-3*NV_DOT(grad_c, A)*gprime; /* flux through face */
}
else /* face lies inside of the domain: use average of adjacent cells. */
{
c1 = F_C1(f,t); /* get cell on other side of face */
t1 = F_C1_THREAD(f,t);
/* current cell gradient of c_eq */
NV_D(grad_c, =, C_UDSI_G(c0,t0,1)[0],C_UDSI_G(c0,t0,1)[1],C_UDSI_G(c0,t0,1)[2]);
/* add value of adjacent cell */
NV_D(grad_c, +=, C_UDSI_G(c1,t1,1)[0],C_UDSI_G(c1,t1,1)[1],C_UDSI_G(c1,t1,1)[2]);
gprime = (gmax + (gzero-gmax)/(1.0+exp(C_UDSI(c0,t0,2)/4.0-342.15/4.0)))/2.0 + (gmax + (gzero-gmax)/(1.0+exp(C_UDSI(c1,t1,2)/4.0-342.15/4.0)))/2.0;
flux = -3E-17/0.988E-3*NV_DOT(grad_c, A)/2.0*gprime;
}
return flux;
}


\end{lstlisting}
The implementation of \autoref{eq:energy} is accomplished with equivalent UDF routines.

\section{Simulation setup}
The global model constants are chosen to match data by Rabeler~\cite{cm_rabeler_mod2018} enabling comparison with simulation data (implemented in COMSOL Multiphysics 5.2a) and corresponding experimental data in validation case \rom{1}. 
We assume a positive\footnote{Rabeler implies a negative value in \cite{cm_rabeler_mod2018}, but this may be rooted in an infelicitous formulation of \autoref{eq:conservation-rtt}. Aggregational effects of a negative diffusion coefficient are not to be expected in the considered physics of diffusive water transport.},
isotropic diffusion coefficient $D = 3\mathrm{e}{-10}$~m$^2$~s$^{-1}$.
The effective material properties for this multiphase scenario are the weighted properties of water and chicken meat (predominantly protein). The volume fraction of each component weights the orthotropic components of the thermal conductivity $\lambda$ 
\begin{align}\label{eq:lambdacp}
\lambda_\text{parallel} &= \lambda_w \phi_w + \lambda_p \phi_p\,,\\
\frac{1}{\lambda_\text{orth.} }&= \frac{\phi_w}{\lambda_w} + \frac{\phi_p}{\lambda_p}\,, 
\end{align}
where $\phi_i$ is defined with the respective volumes $V_j$ as
\begin{align}
\phi_i = \frac{V_i}{\sum_j V_j}.
\end{align}
We assume the fibers of chicken meat to be parallel to the long axis of the cuboid.
The resulting specific heat capacity $c_p$ of chicken meat is weighted with the local concentration 
\begin{align}
c_p = c_{p,w}\, C +c_{p,p}\, (1-C)\,.
\end{align}
Choi and Okos derived analytical temperature-dependent formulae for transport properties of water and protein~\cite{cm_choiokos1986}:
\begin{align} \label{eq:lambdacp-1}
\lambda_w &= 0.57109 + 1.7625\mathrm{e}{-3} \times (T-273.15) - 6.7036\mathrm{e}{-6} \times (T-273.15)^2\,,\\
\lambda_p &= 0.17881 + 1.1958\mathrm{e}{-3} \times (T-273.15) - 2.7178\mathrm{e}{-6} \times(T-273.15)^2\,,\\
c_{p,p} &= 2008.2 + 1.2089\times(T-273.15) - 1.3129\mathrm{e}{-3} \times(T-273.15)^2 \,,\\
c_{p,w} &= 4128.9 - 9.0864\mathrm{e}{-2}\times(T-273.15) + 5.4731\mathrm{e}{-3}\times(T-273.15)^2. \label{eq:lambdacp-4}
\end{align}
%
The global settings for all simulations presented hereafter are denoted in Tab.~\ref{tab:globalsettings}. 
\begin{table*}[hbtp]\centering\footnotesize
\ra{1.05}
\caption{Global simulation settings for Fluent UDS model.}
\vspace*{1em}
\begin{tabular}{@{}ll@{}}\toprule
name & setting / value \\
 \midrule
mesh grid size & $h = 5\mathrm{e}{-4}$ m\\
no. of boundary inflation layers & $n=10$\\
first layer height & $h_1=1\mathrm{e}{-4}$ m\\
spacial discretization & second order upwind\\
time integration & first order implicit Euler \\
time step size & $\Delta t \in [0.1,1]$~s \\
scaled residual threshold & $r < 1\mathrm{e}{-7}$\\
solver type & segregated\\
 \bottomrule
 \label{tab:globalsettings}
\end{tabular}
\end{table*}
The simulations are initialized with constant variable values
\begin{align}
T_0 &= 279.15 \text{ K}\,, \\
C_0 &= 0.76\,.
\end{align}

A grid study with three different mesh sizes has been performed to exclude bias of insufficiently refined grids from the simulation results. The grid convergence index (GCI) is calculated as suggested by ASME Journal of Fluids Engineering Editorial Policy. The method is based on Richardson's fundamental error series expansion and has been refined by Roache~\cite{grid_roache} and Celik~\cite{grid_celik} to be applicable for non-constant refinement factors. 
The measurements of $T_\text{surface}$ after $t = 200$~s are chosen for the grid study.
 Too coarse meshes near the domain surface are expected to produce significant errors. For a mesh with a grid spacing of $h=1.25\mathrm{e}{-4}$~m, the GCI is calculated to be 0.55~\%. The value can be interpreted as a more conservative relative error estimation than it can be provided by traditional Richardson extrapolation. 
 It includes a security factor of $S=1.25$, as recommended for the calculation of the effective convergence order (here $p=0.82$) based on three grid sizes~\cite{grid_celik}. The corresponding absolute error estimation window $E = \pm1.82$~K is added to Fig.~\ref{fig:gci}. 
 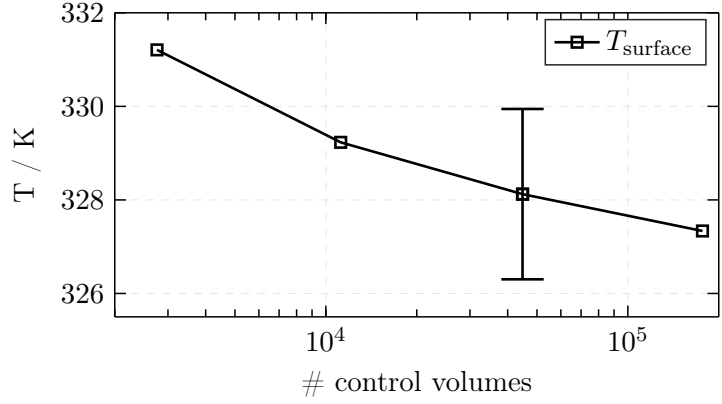
\begin{figure}[htbp]
 \centering
\setlength\figureheightbig{0.23\textwidth}
\setlength\figurewidthbig{0.48\textwidth}	
%
\definecolor{mycolor1}{rgb}{0.00000,0.44700,0.74100}%
\begin{tikzpicture}

\begin{axis}[%
width=0.951\figurewidthbig,
height=\figureheightbig,
at={(0\figurewidthbig,0\figureheightbig)},
scale only axis,
separate axis lines,
every outer x axis line/.append style={black},
every x tick label/.append style={font=\color{black}},
every x tick/.append style={black},
xmode=log,
xmin=2000,
xmax=200000,
xminorticks=true,
xlabel={\# control volumes},
every outer y axis line/.append style={black},
every y tick label/.append style={font=\color{black}},
every y tick/.append style={black},
ymin=325.5,
ymax=332,
ylabel={T / K},
axis background/.style={fill=white},
xmajorgrids,
ymajorgrids,
grid style={line width=.1pt, draw=gray!10},
major grid style={line width=.2pt,draw=gray!20}, 
legend style={legend cell align=left, align=left, draw=white!15!black}
]
\addplot [color=black, line width=1pt,  mark=square, mark options={solid, black}]
  table[row sep=crcr]{%
2760 	331.207169\\
11200	329.23090\\
44800	328.123347\\
176640 327.3358951064525 \\
};

\addlegendentry{$T_\text{surface}$}

\addplot [color=black, line width=1pt, mark=square, mark options={solid, black}]
 plot [error bars/.cd, y dir = both, y explicit,error bar style={line width=1pt},
     error mark options={
      rotate=90,
      black,
      mark size=8pt,
      line width=1pt
    }]
 table[row sep=crcr, y error plus index=2, y error minus index=3]{%
0	326.666701292857	1.82080713392898	1.82080713392898\\
44800	328.123347	1.82080713392898	1.82080713392898\\
};

\end{axis}

\end{tikzpicture}%
 \caption{Grid convergence study for $T_\text{surface}$ at t = 200 s.}
 \label{fig:gci}
\end{figure}
Course meshes tend to produce simulation results that over-predict the temperatures. An additional simulation with $h=6.25\mathrm{e}{-5}$~m confirms this assumption. Hence, the error window can be narrowed to $E = -1.82$~K.
As steep gradients are expected, particularly below the surfaces of the cuboid, mesh inflation with a sufficiently small initial grid spacing $h_1$ is applied. 

\section{Case \rom{1}: Chicken cuboid in a convection oven}\label{sec:case1}

The coupled C-T model is validated with the benchmark case of a chicken cuboid on a baking plate inside a convection oven. 
A quarter sector model of the cuboid is set up (compare Fig.~\ref{fig:cube-measured}). Position A marks the center of the cuboid where the temperature $T_\text{core}$ is measured. Position B lies above position A and one millimeter below the surface. Here, $T_\text{surface}$ is probed.
\begin{figure}[H]
 \centering
 \includegraphics[width=0.33\textwidth]{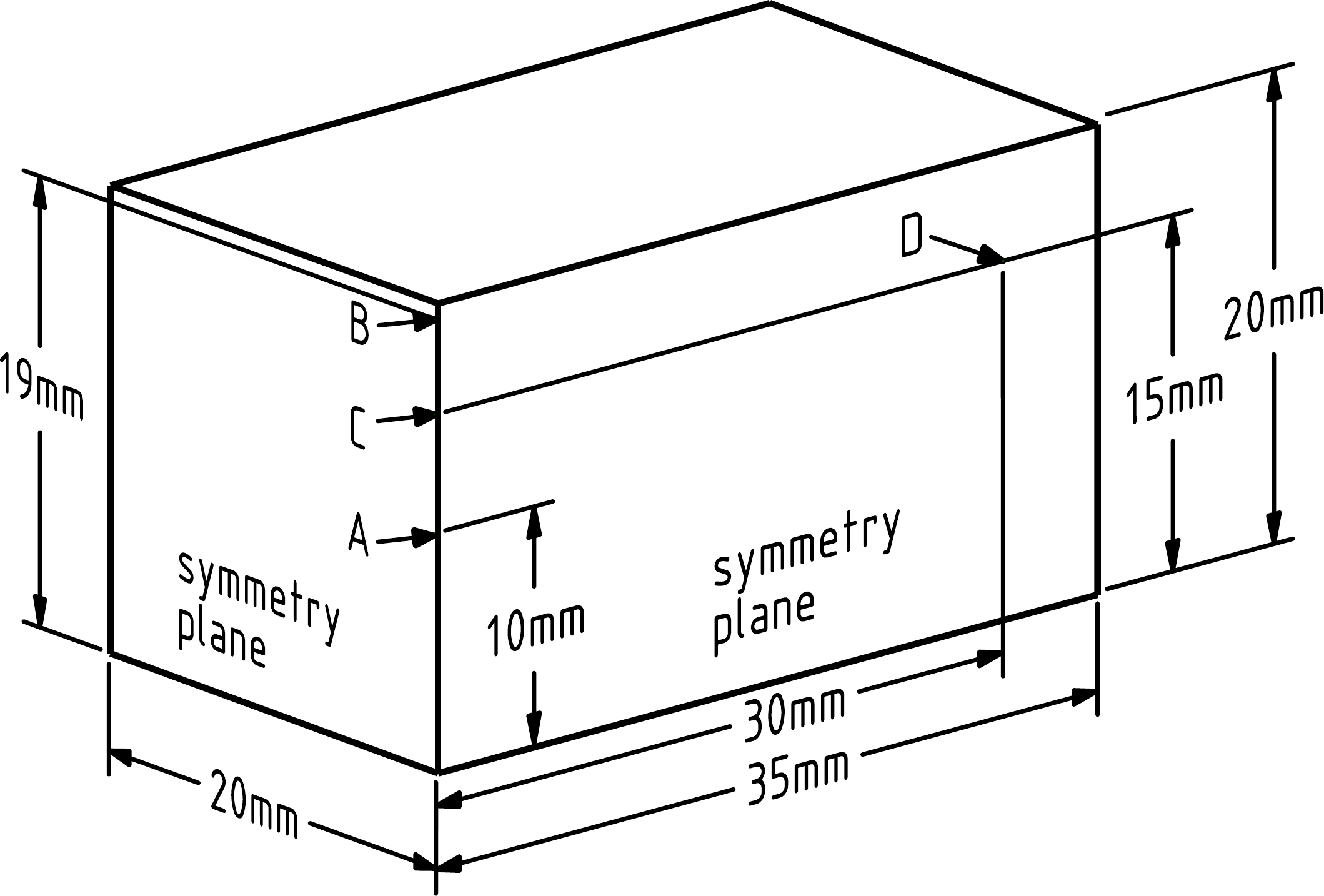}
 \caption{Model setup of case \rom{1} with measurement positions for $T_\text{core}$ and $T_\text{surface}$.}
 \label{fig:cube-measured}
\end{figure}
The Neumann-type boundary conditions for case \rom{1} are denoted in Tab.~\ref{tab:bc-conv}. Due to the direct contact of the chicken meat with the baking plate, we see higher heat transfer coefficients at the bottom of the cuboid. The ambient temperature in the convection oven is set to $T_\text{oven} = 443.15 \text{ K}$ and the concentration of water in hot air is five per cent. 
\begin{table*}[hbtp]\centering\footnotesize
\ra{1.3}
\caption{Boundary conditions for case \rom{1}: cuboid in a convection oven.}
\vspace*{1em}
\footnotesize
\addtolength{\tabcolsep}{-3pt}
\begin{tabular}{@{}llrlr@{}}\toprule
position & C boundary condition & & T boundary condition &
\\ \midrule 
bottom & $ \left( D_{ik} \frac{\partial C}{\partial x_k} - C\,u_i \right)n_i = -\beta_\text{bot}\left(C-0.05\right) $, & $\beta_\text{bot}=1\mathrm{e}{-6} $m s$^{-1}$
& $\lambda\frac{\partial T}{\partial x_i}n_i = -\alpha_\text{bot} \left(T - T_\text{oven}\right)$, & $\alpha_\text{bot}$ = 59 W m$^{-2}$ K$^{-1}$ \\
surface & $ \left( D_{ik} \frac{\partial C}{\partial x_k} - C\,u_i \right)n_i = -\beta_\text{surf}\left(C-0.05\right), $ & $\beta_\text{surf}=1\mathrm{e}{-6} $m s$^{-1}$ 
& $\lambda\frac{\partial T}{\partial x_i}n_i = -\alpha_\text{surf} \left(T - T_\text{oven}\right)$, & $\alpha_\text{surf}$ = 44 W m$^{-2}$ K$^{-1}$ \\
symmetry & $ \left( D_{ik} \frac{\partial C}{\partial x_k} - C\,u_i \right)n_i = 0 $ m s$^{-1}$ &
& $\lambda\frac{\partial T}{\partial x_i}n_i = 0$ / W m$^{-2}$ & \\ 
 \bottomrule 
 \label{tab:bc-conv}
\end{tabular}
\end{table*}
While cooking chicken meat, its fiber structure contracts and water is expelled. 
The convection term $u_i$ transports water from the inside towards the surface of the chicken (compare Fig.~\ref{fig:sim-uw}). The contribution of the term $\rho\, c_p\, u_i$ in~\autoref{eq:energy} affects a supplementary cooling of the surface region. Temperatures of the surface and in the core would be up to ten Kelvin higher without the simulation of this coupled transport effect (see Fig.~\ref{fig:sim-modelling-coupled}). Besides the predominantly outward-facing flow of water, we can distinguish a transport of low magnitude $u_i \approx 1\mathrm{e}{-8}$~m~s$^{-1}$ towards the core of the cuboid. Van der Sman has observed this behavior as well~\cite{cm_vandersman_moisture2007a}. Its low magnitude coincides with statements by Feyissa, who only noticed a weak inward-facing effect in his simulations~\cite{cm_feyissa_3D2013}.
\begin{figure}[htbp]\small
\centering
\centering
 \includegraphics[width=0.65\textwidth]{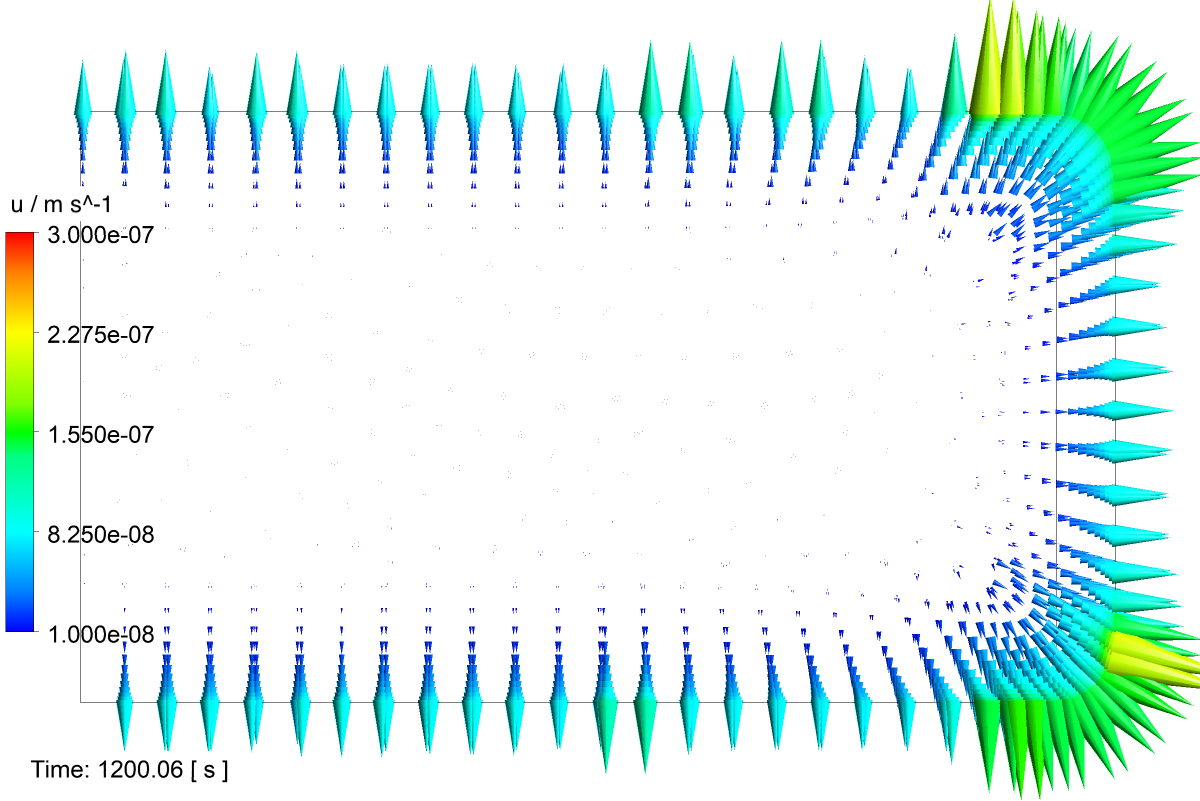}
 \caption{Influence of the water convection $u_i$ in the surface region -- view on the long-sided symmetry plane.}
 \label{fig:sim-uw}
\end{figure}
Both reference si\-mu\-la\-tion and the implementation of the C-T-model in ANSYS Fluent show a good agreement with experimental data. A water loss to an average water concentration of 62~\% after 1200 seconds is reproduced. A closer comparison of the temperature trends at core and surface position reveals that both implementations operate in a similar error range compared to the experimental data, see Fig.~\ref{fig:sim-error-cube}. However, the implemented Fluent model reduces the root-mean-square and maximum error, partly by factor two (see Tab.~\ref{tab:error-case1}), although simple Neumann boundary conditions for $C$ have been applied, compare Tab.~\ref{tab:bc-conv}. 
The overshoot of $T_\text{surface}$ during the first 200 seconds of the reference simulation could presumably be related to insufficiently refined meshes in the boundary layer.
\begin{figure}[htbp]\small
\captionsetup[subfigure]{justification=centering}
\centering
\begin{subfigure}[c]{0.49\textwidth}
\setlength\figureheightbig{0.3\textheight}
\setlength\figurewidthbig{0.95\textwidth}	
%
%
\definecolor{mycolor1}{rgb}{0.00000,0.44700,0.74100}%
\definecolor{mycolor2}{rgb}{0.85000,0.32500,0.09800}%
\definecolor{mycolor3}{rgb}{0.92900,0.69400,0.12500}%
\definecolor{mycolor4}{rgb}{0.49400,0.18400,0.55600}%
\begin{tikzpicture}
\begin{axis}[%
width=0.951\figurewidthbig,
height=\figureheightbig,
xmin=0,
xmax=1200,
xlabel={t / s},
ymin=270,
ymax=380,
ylabel={T / K},
axis background/.style={fill=white},
legend style={at={(0.97,0.03)}, anchor=south east, legend cell align=left, align=left, draw=white!15!black}
]
\addplot [color=black, only marks, mark=square, mark options={solid, black}]
  table[row sep=crcr]{%
4.65211601399393	279.880361509235\\
58.1198961742468	280.897101531054\\
119.057036451868	284.066946920964\\
178.900282135199	290.27565173231\\
237.666899898431	297.37063724937\\
299.632930820449	305.605448971147\\
359.457894142873	314.093354775608\\
418.22654327954	320.935095737878\\
479.137275702517	327.397120340067\\
537.91405033292	333.225883083173\\
598.830876876199	338.928174020991\\
656.543211827108	344.123750141068\\
719.608215777001	348.686591054433\\
777.33070759508	352.615944400557\\
838.26480081255	356.165656622654\\
899.198894030019	359.71536884475\\
959.075657375014	361.74553850205\\
1018.95242072001	363.77570815935\\
1079.8997178648	365.679330775308\\
1138.71305721702	366.949691532182\\
1197.52842794267	367.966807734266\\
};
\addlegendentry{$T_\text{core}$ Experiment}

\addplot [color=black,line width=1pt]
  table[row sep=crcr]{%
0	279.15\\
15	279.15004960213\\
30	279.158984587679\\
45	279.246815484258\\
60	279.527786789587\\
75	280.076075890156\\
90	280.908313337782\\
105	282.005289257995\\
120	283.332604049583\\
135	284.852194161869\\
150	286.527469872022\\
165	288.325332876833\\
180	290.21684310969\\
195	292.177241026086\\
210	294.185646131406\\
225	296.224640758342\\
240	298.279857741424\\
255	300.339608403261\\
270	302.394544045778\\
285	304.437338445846\\
300	306.462386983251\\
315	308.465524901536\\
330	310.443769704963\\
345	312.395092739575\\
360	314.318219175537\\
375	316.212454335197\\
390	318.077531927045\\
405	319.913482030539\\
420	321.72051871463\\
435	323.498946165199\\
450	325.249081770922\\
465	326.971193435691\\
480	328.665445304033\\
495	330.331843976883\\
510	331.970176450387\\
525	333.579935420842\\
540	335.160264179243\\
555	336.709982179199\\
570	338.227757352305\\
585	339.712435358084\\
600	341.163430346594\\
615	342.581000816035\\
630	343.965260315799\\
645	345.311221930831\\
660	346.610897741651\\
675	347.860462314634\\
690	349.060258256211\\
705	350.211990038475\\
720	351.31679506899\\
735	352.37488998216\\
750	353.386129586329\\
765	354.350620112121\\
780	355.268773745702\\
795	356.141523445503\\
810	356.970230101736\\
825	357.756436334805\\
840	358.501832714222\\
855	359.208195301995\\
870	359.877333484608\\
885	360.511051031686\\
900	361.11111859045\\
915	361.679255209454\\
930	362.217116731782\\
945	362.7262891553\\
960	363.208285459292\\
975	363.664544894979\\
990	364.096433922609\\
1005	364.505248217424\\
1020	364.892215292125\\
1035	365.258497492555\\
1050	365.605195156902\\
1065	365.933349796581\\
1080	366.243948493688\\
1095	366.537925940441\\
1110	366.816162976031\\
1125	367.079495183518\\
1140	367.328714288016\\
1155	367.564570122431\\
1170	367.78777279822\\
1185	367.998994824066\\
1200	368.198873111585\\
};
\addlegendentry{$T_\text{core}$ Fluent}

\addplot [dashed,color=black, line width=1.0pt]
  table[row sep=crcr]{%
13.643429732933	279.096341091078\\
40.4835157205602	279.517243906435\\
68.6057076469447	280.64677009926\\
96.7236346864374	282.307985524604\\
125.479650524172	284.223522938694\\
152.94596943502	287.315826232214\\
181.053844299212	290.23021284502\\
209.160170677155	293.33764407691\\
236.382061334518	296.719485471688\\
263.527339964468	299.653634495939\\
289.788561563167	302.814983746955\\
317.893609078564	306.081846509472\\
345.798430802643	309.396867273314\\
371.610690258143	312.767494810985\\
399.904322608819	315.706488208965\\
426.802818203156	319.025716521872\\
456.116430499852	321.989277157749\\
484.223055827414	325.059439337038\\
512.331090741803	327.95387302539\\
540.43924263282	330.833723627326\\
568.548910011613	333.524643419887\\
596.658624753639	336.209658596128\\
624.7702057563	338.662013279826\\
652.70565841266	341.120101649257\\
680.994972250621	343.366696351967\\
709.10831400774	345.599543646496\\
737.221995665619	347.790016646362\\
765.338030473134	349.687130324936\\
793.453729202651	351.626141727251\\
821.570247667967	353.462959399916\\
849.686736115896	355.303519240218\\
877.805522852471	356.857559095945\\
905.92390196591	358.46241596936\\
934.043978402293	359.855673248977\\
962.16351011849	361.316838978479\\
990.284610878975	362.582397186847\\
1018.40626515969	363.778949873135\\
1046.52767724303	365.005696498817\\
1074.65118339174	365.971382975665\\
1102.77381740094	367.045796176664\\
1130.89781105893	367.950706496298\\
1159.02132702152	368.915169508252\\
1187.14631490075	369.69613357966\\
1202.36183302503	370.073649215332\\
};
\addlegendentry{$T_\text{core}$ Reference}

\end{axis}

\end{tikzpicture}%
\subcaption{Core temperatures.}
\end{subfigure}
\begin{subfigure}[c]{0.49\textwidth}
\setlength\figureheightbig{0.3\textheight}
\setlength\figurewidthbig{0.95\textwidth}	
\input{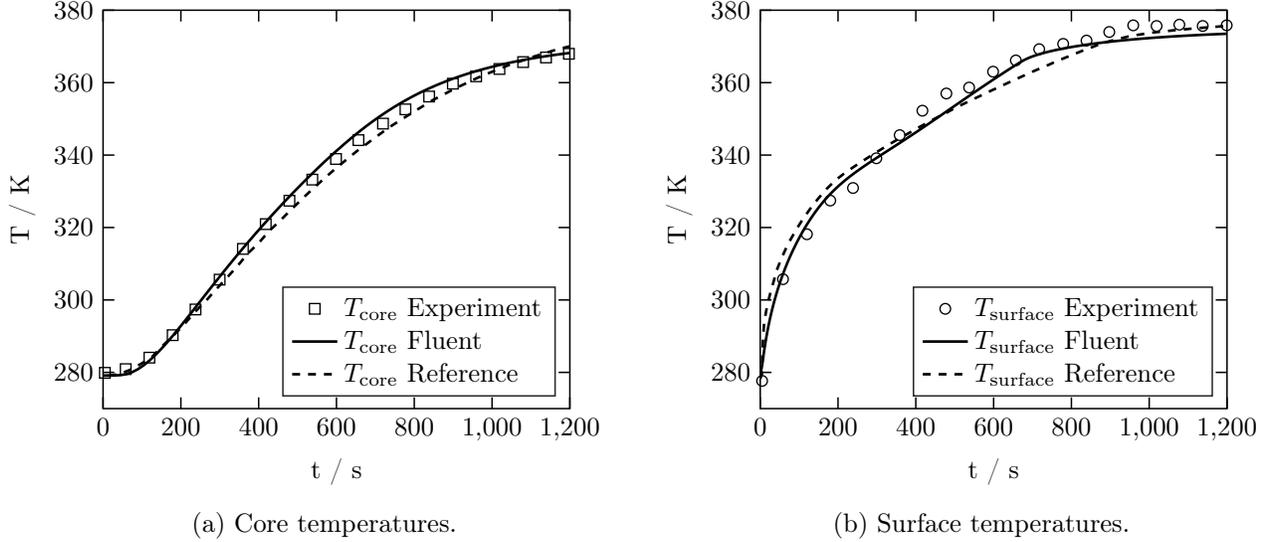} 
\subcaption{Surface temperatures.}
\end{subfigure}
 \caption{Validation of temperatures with experimental data and reference simulation by Rabeler et al. \cite{cm_rabeler_mod2018}.}
 \label{fig:sim-error-cube}
\end{figure}
\begin{table*}[hbtp]\centering\footnotesize
\ra{1.1}
\caption{Error calculation for case \rom{1}.}
\vspace*{1em}
\begin{tabular}{@{}lllll@{}}\toprule
position & $E_\text{RMS}$ Fluent & $E_\text{RMS}$ reference & $E_\text{max}$ Fluent & $E_\text{max} $ reference \\
 \midrule
$T_\text{surface}$ & 1.87 K & 3.19 K & 4.27 K & 12.03 K\\
$T_\text{core} $ & 1.43 K & 2.78 K & 2.61 K & 5.47 K\\
 \bottomrule
 \label{tab:error-case1}
\end{tabular}
\end{table*}

Water's evaporation at the cube's surface hinders temperatures from rising considerably beyond the boiling point. Feyissa proposes a smooth reduction of energy input~\cite{cm_feyissa_3D2013}. Based on estimations of the residual latent heating of the probe, he suggests a throttling to approximately 10~\% of the original heat flux. These modified Neumann boundary conditions introduce nonlinearity in the model. The reduced heat throughput beyond boiling temperatures must be learned during the reduced-order model's training phase, see Section~\ref{sec:rom}. The absence of an evaporation model implies steadily rising temperatures of the chicken, which do not agree with experimental data (see Fig.~\ref{fig:sim-modelling-evap}).

\begin{figure}[tbhp]
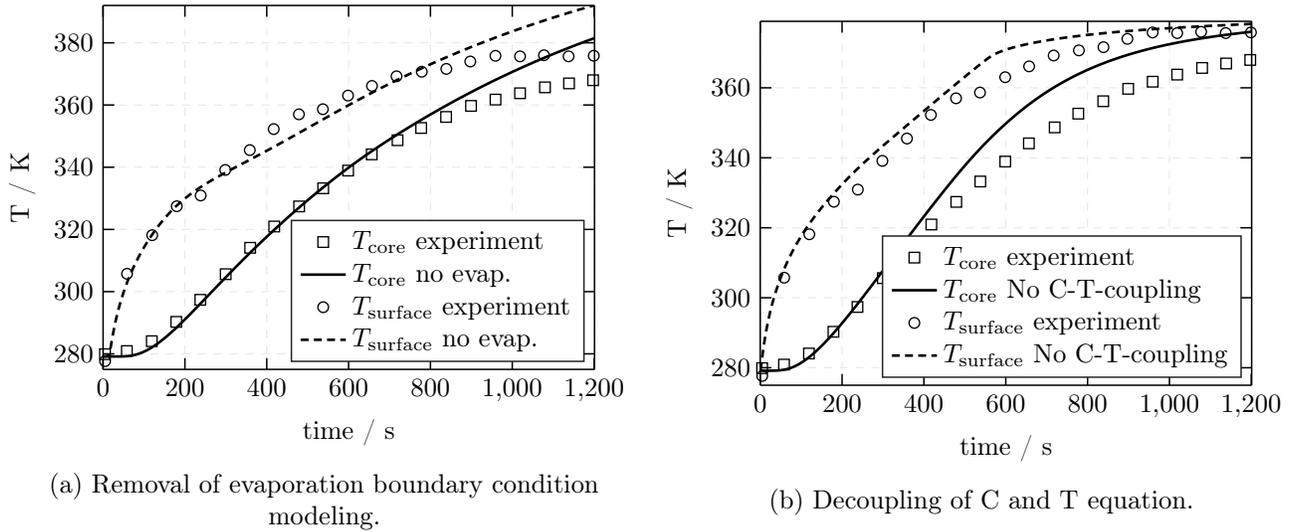
\small
\captionsetup[subfigure]{justification=centering}
\centering
\begin{subfigure}[c]{0.49\textwidth}
\setlength\figureheightbig{0.28\textheight}
\setlength\figurewidthbig{0.99\textwidth}
\input{tiks/garmod/T_ab_modelling2.tex} 
 \subcaption{Removal of evaporation boundary condition modeling.}
 \label{fig:sim-modelling-evap}
\end{subfigure}
\begin{subfigure}[c]{0.49\textwidth}
\setlength\figureheightbig{0.28\textheight}
\setlength\figurewidthbig{0.99\textwidth}
\input{tiks/garmod/T_ab_modelling_nocoupl.tex} 
 \subcaption{Decoupling of C and T equation.}
 \label{fig:sim-modelling-coupled}
\end{subfigure}
 \caption{Effects of model changes to the core and surface temperature trends.}
 \label{fig:sim-modelling}
\end{figure}

\subsection{Case \rom{2}: Pan-frying chicken meat}
This validation case verifies the applicability of the implemented C-T-model to pan-frying of chicken meat. 
A thermal fluid-structure interaction simulation is set up to estimate the heat transfer coefficient to ambient air. Nine realistically shaped chicken fillets in a mid-sized industry-scale pan-frying device are modeled in {ANSYS} SpaceClaim 2019R2 and implemented in ANSYS CFX 2019R2, see Fig.~\ref{fig:vccacad_setup}. 
\begin{figure}[bhtp]\small
\captionsetup[subfigure]{justification=centering}
\centering
\begin{subfigure}[c]{0.4\textwidth}
\centering
\includegraphics[width=0.25\textheight]{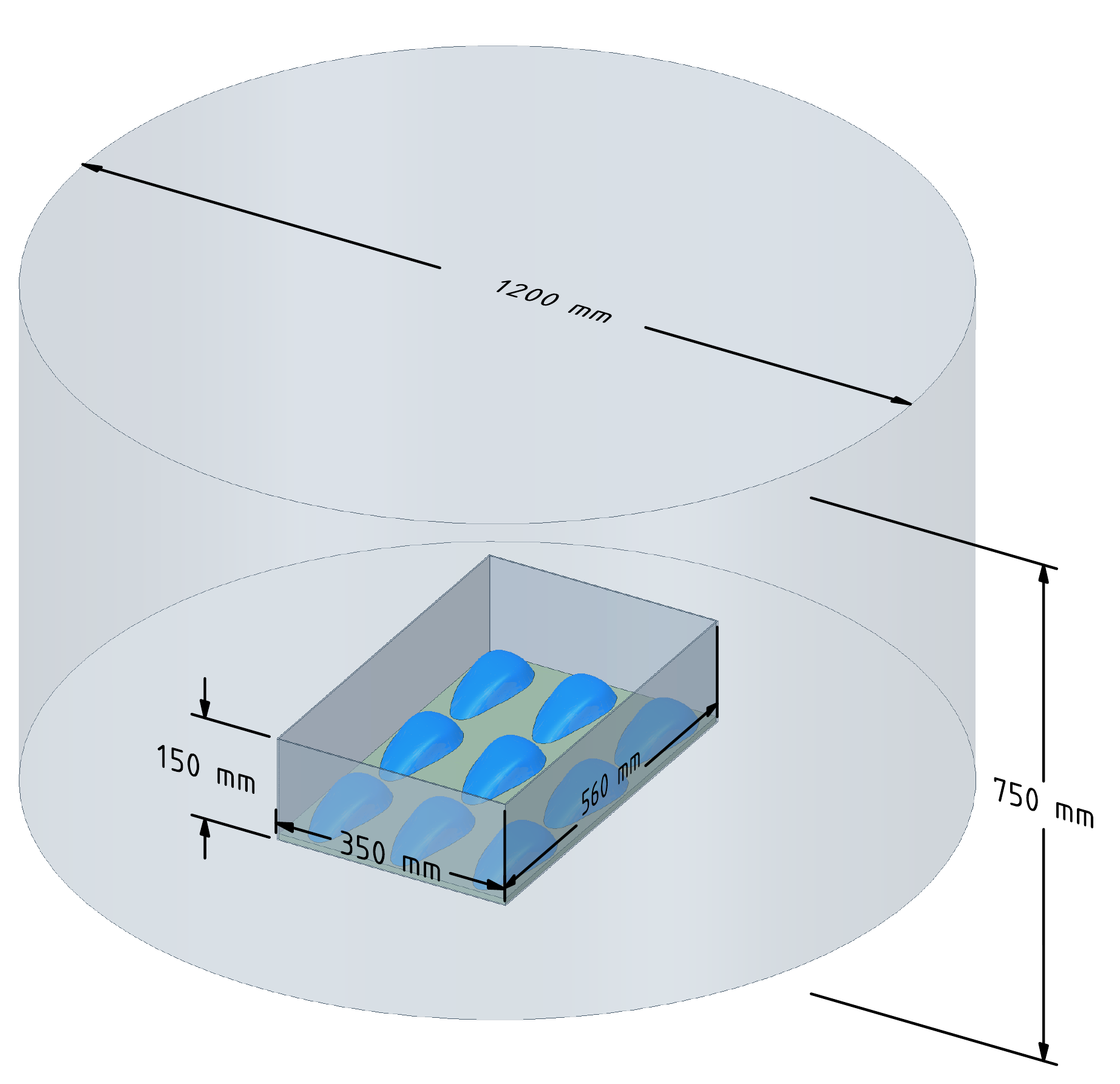}
\subcaption{Geometric dimensions.}
\end{subfigure}
\begin{subfigure}[c]{0.48\textwidth}
\centering
\includegraphics[height=0.25\textheight]{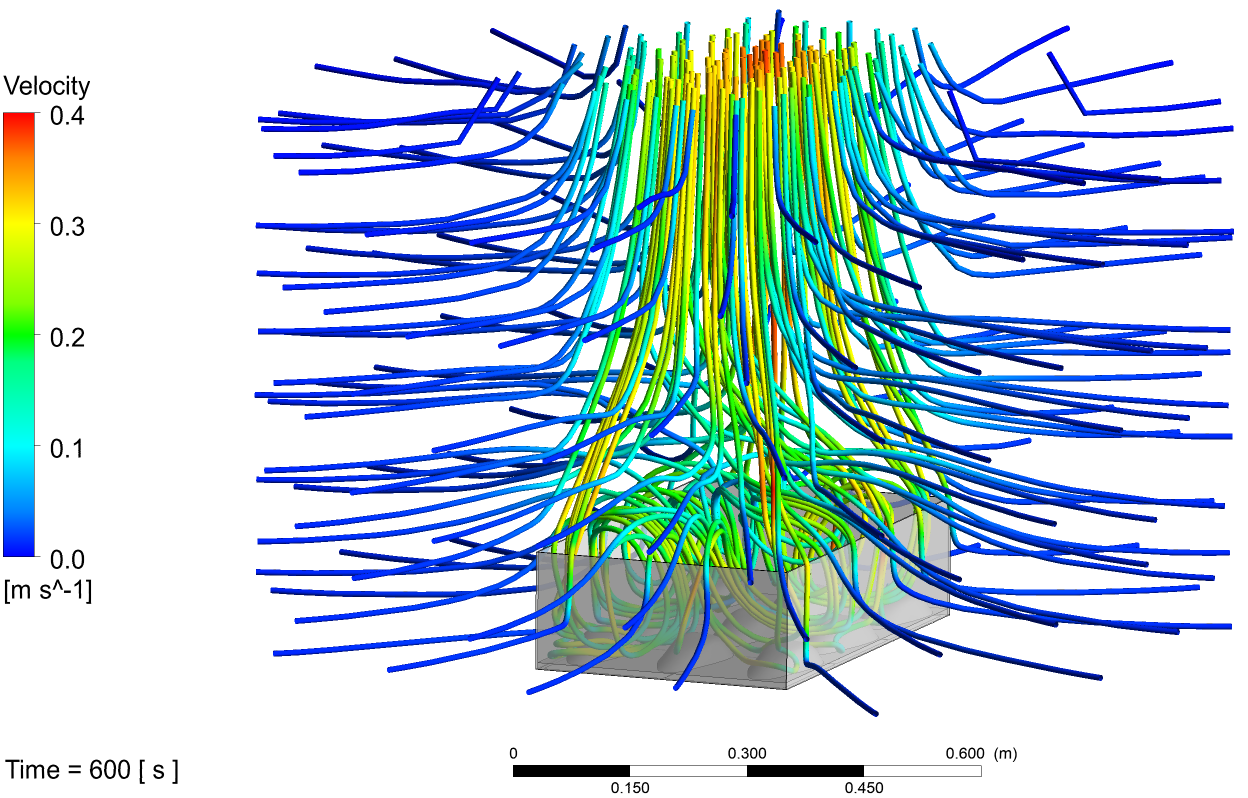}
\subcaption{Natural convection streamlines.}
\label{fig:vccacad_setup_b}
\end{subfigure}
 \caption{TFSI simulation setup of case \rom{2} incl. radiation and natural convection modelling.}
 \label{fig:vccacad_setup}
\end{figure}
The material properties of steel (grade~304) and air are modeled temperature dependent with data of VDI Wärmeatlas~\cite{thermo_waermeatlas}. The chicken material properties match ~\autoref{eq:lambdacp-1} to~\autoref{eq:lambdacp-4} with a protein content of 22~\%.
Recent thermodynamic studies of industry-grade cooking devices~\cite{mscthesis} show that radiation effects cannot be neglected at temperature ranges of pan-frying. The grey discrete transfer radiation model with participating media effects (interaction with air) is chosen. Based on the experimental emissivity measurements by Ibarra et al.~\cite{cm_ibarra_emissivity2000}, we choose an average emissivity of $\epsilon=0.8$ for chicken meat. 
The realization of natural convection in CFX needs special attention. The full buoyancy source term reads
\begin{align}
S_k=\left(\rho - \rho_\text{ref}\right) g_k\,.
\end{align}
The local density of air can be calculated with the ideal gas law
\begin{equation}
\rho = \frac{M \,p_\text{abs} }{R \,T} = \frac{M \left[ p_\text{ref}+p_\text{stat} +\rho_\text{ref} \, g_i \left(r_i(x_k) - r_{i\text{,ref}}\right) \right] }{R\, T }\,,
\end{equation}
where the absolute pressure compromises a reference pressure $p_\text{ref}$ at a reference position $r_{i\text{,ref}}$, the static pressure $p_\text{stat}$ and a hydrostatic pressure component. For domains with ``Opening'' boundaries present, compare Fig.~\ref{fig:vccacad_setup}, the hydrostatic pressure gradient would induce buoyancy forces. Consequently, the flow would self-accelerate until the forces are in equilibrium with the viscous forces. As magnitudes of up to $u_\text{buoy} \approx 0.16$~m~s$^{-1}$ can be reached for the presented case, the resulting error of $40\%$ is prevented with a custom ideal gas law, which neglects the hydrostatic pressure term. Consequently, the streamlines in Fig.~\ref{fig:vccacad_setup_b} enter the fluid domain horizontally, and fluid flow only rises vertically due to buoyancy induced by temperature gradients.
For each chicken, the total heat flux $q_\text{total}$ is averaged at the boundaries facing ambient air. 
The average external heat transfer coefficient is found to be 
\begin{align}
\alpha_\text{air} = \frac{q_\text{total}}{|T-T_\text{ambient}|} \approx 15~W~m^{-2}~K^{-1}.\end{align}

\subsubsection{Comparison with experimental results}

Several conventional chicken breast fillets with an initial temperature of $T_0 = 280.0 \pm 2$~K have been pan-fried by an industrial project partner.
\begin{figure}[htbp]\small
\captionsetup[subfigure]{justification=centering}
\centering
\begin{subfigure}[c]{0.48\textwidth}
\centering
\includegraphics[height=0.1\textheight]{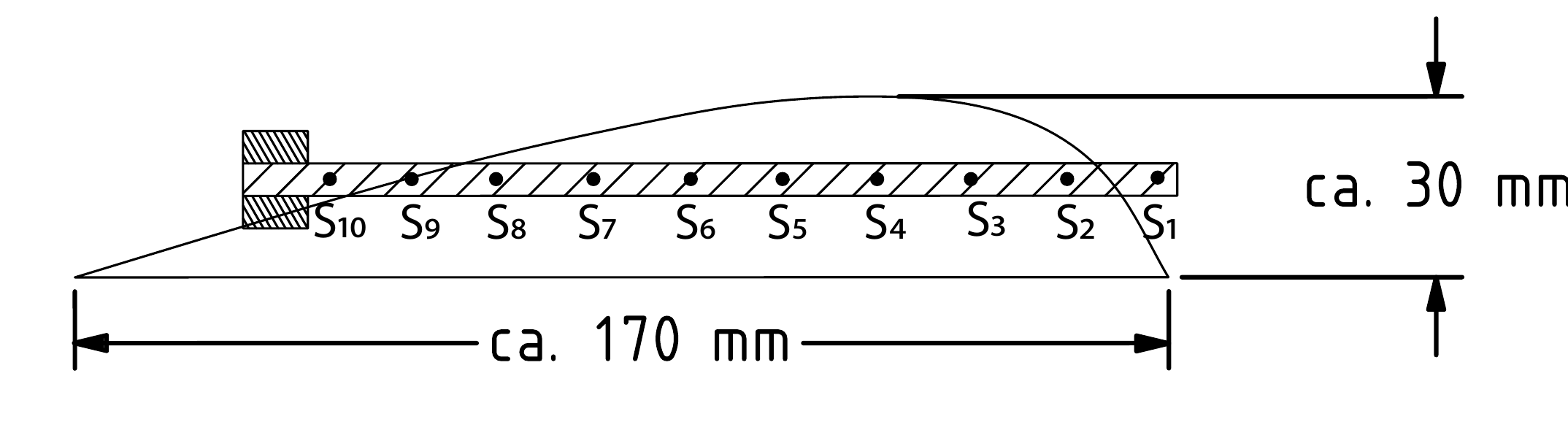} 
\subcaption{Measurement positions for $S_1$ to $S_{10}$.}
\label{fig:measpos}
\end{subfigure}
\begin{subfigure}[c]{0.48\textwidth}
\centering
\setlength\figureheightbig{0.23\textheight}
\setlength\figurewidthbig{0.99\textwidth}	
\input{tiks/garmod/VCC_vs_Experiment.tex} 
 \caption{Temperature corridor of simulated and experimentally obtained core temperatures during pan-frying.}
 \label{fig:corridorpanfry}
\end{subfigure}
 \caption{Experimental setup and results for case \rom{2}.}
 \label{fig:exp-KTchickens}
\end{figure}
The temperature of the contact heating device is set to $T=443.15$~K. Additional thermocouples confirm that the pan temperatures are $T=443.15 \pm 3$~K throughout the experiments. 
Core temperatures are recorded with a pointed tube probe containing multiple thermocouples (see Fig.~\ref{fig:measpos}). The probe is inserted mid-height into the chicken at approximately 15~mm above the bottom level. As there is an inevitable spread in the correct positioning of the sensor and each fillet has a natural deviation, we only use the data $S_4$ and $S_5$ for comparison. The virtual probe points $C$ and $D$ are positioned 15~mm above the bottom level and inside the cuboid's long symmetry plane (see Fig~\ref{fig:cube-measured}). Considering the spread of initial temperatures, positioning of the sensor and the geometric deviations of chicken fillet samples, we see a good agreement of simulations and experiments in Fig.~\ref{fig:corridorpanfry}. The simulation data shows a maximum deviation of $E_\text{pan-fry, max}=+5.42$~K after 900~seconds. During the first ten minutes, the errors remain in the bounds of $E_\text{pan-fry, max}=\pm 2$~K.
%

\section{Setup of a Digital Twin}\label{sec:rom}
The core framework of a Digital Twin is usually a closed-loop control algorithm. It receives live measurement data of the considered process. In this case, it could be the ambient temperature, the de-facto pan temperature or natural convection conditions, e.g., the state of a ventilation system. The Digital Twin then predicts the future trend of the process quantities ($T_\text{core}$ or the residual water content at the end of the cooking process). The feedback system reacts to the newly obtained information on the process and adjusts the input quantities accordingly (such as the pan temperature $T_\text{input}$). The discrete refresh times of the control loop should be in the order of seconds to assure fast dynamic behavior. CFD co-simulations become unfeasible due to several reasons. The calculation of one hour of real time in the presented C-T model requires more than one hour of computational time on a 16-core Intel Xeon CPU with a 3.2~GHz base clock. Hence, a real-time prediction of the end-state is impossible during the control phase. Besides, even state-of-the-art cooking devices would not be equipped with a professional multicore CPU. 

Reduced-order models (ROM) are derived to circumvent the downsides of co-simulation. The evaluation of a small-sized ODE system~\cite{rom_xiao11} can be accomplished in milliseconds on the machine mentioned above. We can differentiate ROMs that are intrusive or non-intrusive. The latter are stand-alone mathematical methods that only rely on transient solution data of input and output~\cite{ansys_twinbuilder_2019R2}, and they do not access the physics of the full-order model like their intrusive counterparts. Furthermore, we can introduce a classification of linear or non-linear systems. 
We will investigate the capabilities of ANSYS Twin Builder's non-intrusive toolkits for reduced-order modeling that are offered in the ANSYS Release 2019R2. 

\subsection{LTI and dynamic reduced-order modeling}
The training phase of ROMs needs appropriate training data of the considered model.
Following the system dynamics theory, a system's behavior can be fully characterized by its step response. The step response of the C-T model is recorded. The pan temperature is set to $T_\text{in}=443.15$~K while the core temperature $T_C$ is recorded over time.

\begin{figure*}[htbp]\small
\captionsetup[subfigure]{justification=centering}
\centering
\begin{subfigure}[c]{0.3\textwidth}
\setlength\figureheightbig{0.25\textheight}
\setlength\figurewidthbig{0.99\textwidth}	
\includegraphics[width=0.92\textwidth]{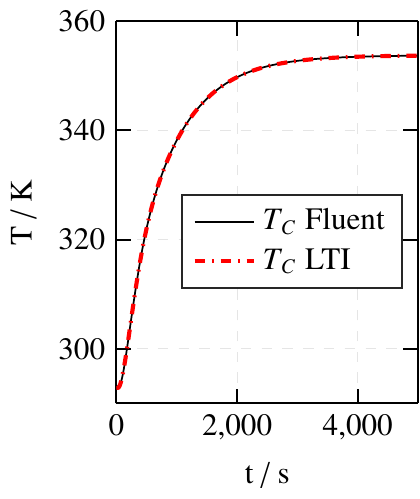}
\subcaption{Constant $T=443.15$~K.}
 \label{fig:rom-dyn-allerror-a}
\end{subfigure}
\begin{subfigure}[c]{0.3\textwidth}
\setlength\figureheightbig{0.25\textheight}
\setlength\figurewidthbig{0.99\textwidth}	
\input{tiks/garmod/ROM_vs_LTI_Const473.tex} 
\subcaption{Constant $T=473.15$~K.}
 \label{fig:rom-dyn-allerror-b}
\end{subfigure}
\begin{subfigure}[c]{0.3\textwidth}
\setlength\figureheightbig{0.25\textheight}
\setlength\figurewidthbig{0.99\textwidth}	
\input{tiks/garmod/ROM_vs_LTI_Rect2.tex} 
\subcaption{Rectangular Pulse $T=443.15$~K.}
 \label{fig:rom-dyn-allerror-c}
\end{subfigure}
 \caption{Comparison of LTI ROM, Dynamic ROM and Fluent data.}
 \label{fig:rom-dyn-allerror}
\end{figure*}

State-space models provide a representation of dynamic systems that are linear and time-invariant (LTI). Twin Builder offers the ``thermal model identification'' toolkit to generate LTI ROMs. The underlying vector fitting algorithm performs the curve fit of a low-dimensional ODE system to the transient input and output signal in the frequency domain~\cite{rom_xiao11}. A relative error of 0.5~\% is set during the fitting procedure. The LTI ROM can reproduce the trained step response without significant error, see Fig.~\ref{fig:rom-dyn-allerror-a}. In Section~\ref{sec:case1}, we discussed the inherent non-linearity of evaporation effects at boiling temperatures. As expected, the linear ROM fails to predict the simulation data of a step response to $T_\text{in} = 473.15$~K, compare Fig.~\ref{fig:rom-dyn-allerror-b}. It consequently over-predicts the core temperature by an error of $3.6$~K. As this error may still be tolerable for some applications, we conclude the investigation of LTI ROMs with a rectangular impulse input with a magnitude of $T_\text{in}=443.15$~K for 60~s. The LTI ROM fails to predict the cooling phase of the chicken fillet, see Fig.~\ref{fig:rom-dyn-allerror-c}. Hence, LTI ROMs are not suitable for cooking use cases.

Since the release of 2019R1, ANSYS Twin Builder has offered the Dynamic ROM Builder toolkit for ROM creation. It allows non-linear physics and multiple transient training data. Traditionally, ANSYS remains predominantly silent on the theory and algorithms of the tools besides a small comment that it can be considered an ``adaptation of the classical linear approach to cope with a quadratic structure of equations [...]''. Non-linear system behavior is ``learned from the data thanks to machine learning approaches''~\cite{ansys_twinbuilder_2019R2}.
We provide four histories of learning data, see~Tab.~\ref{tab:trainingsettings}. 
\begin{table*}[hbtp]\centering\footnotesize
\ra{1.05}
\caption{Learning and evaluation phase excitation signals for $T_\text{in}$.}
\vspace*{1em}
\begin{tabular}{@{}llll@{}}\toprule
& case & description & $E_\text{max}$ \\
 \midrule
\multirow{ 5}{*}{learning phase} & $T=443.15$~K & Constant input with $t_\text{up}=10$ s ramp up & $0.16$~K\\
 & $T=473.15$~K & Constant input with $t_\text{up}=10$ s ramp up & $0.12$~K\\
& $T=403.15$~K & Trapezoidal pulse, $t_\text{up}=200$ s,$t_\text{const}=300$ s, $t_\text{down}=100$ s & $0.13$~K\\
 & $T=443.15$~K & Sawtooth, $t_\text{period}=500$ s & $0.19$~K\\
 & $T=498.15$~K & Half-sine pulse, $t_\text{period}=500$ s & $0.15$~K\\
 \midrule
\multirow{ 4}{*}{evaluation phase} & $T=443.15$~K & Rectangular pulse, $t_\text{pulse}=60$ s& $0.32$~K\\
 & $T=443.15$~K & Half-sine pulse, $t_\text{pulse}=200$ s& $0.22$~K\\
 & $T=443.15$~K & Constant input, no ramp up & $0.12$~K\\
 & $T=473.15$~K & Three Half-sine pulses, $t_\text{pulse}=50$ s, $t_\text{dead}=450$ s& $0.44$~K\\
 \bottomrule
 \label{tab:trainingsettings}
\end{tabular}
\end{table*}
We cover the typical range of pan temperatures for frying and use smooth (sine) and non-smooth (sawtooth etc.) signals for excitation. After the learning phase, we evaluate the quality of the ROM with evaluation signals that have not been used during learning. The absolute error values are less than 0.2~K for trained excitations and less than 0.45~K for untrained signals (see Tab.~\ref{tab:trainingsettings}). The time data for the worst-case scenarios of the learning and evaluation phase is depicted in Fig.~\ref{fig:rom-dyn-examples}. The predominant error source is high-frequency oscillations at the onset of a signal change. However, the sawtooth learning case demonstrates that the error is hardly noticeable at higher operational temperatures. All in all, the temperature error is less than the model error or the discretization error of the full-order model by factor ten and is therefore uncritical.
\begin{figure*}[htbp]\small
\captionsetup[subfigure]{justification=centering}
\centering
\begin{subfigure}[c]{0.48\textwidth}
\centering
\includegraphics[width=0.92\textwidth]{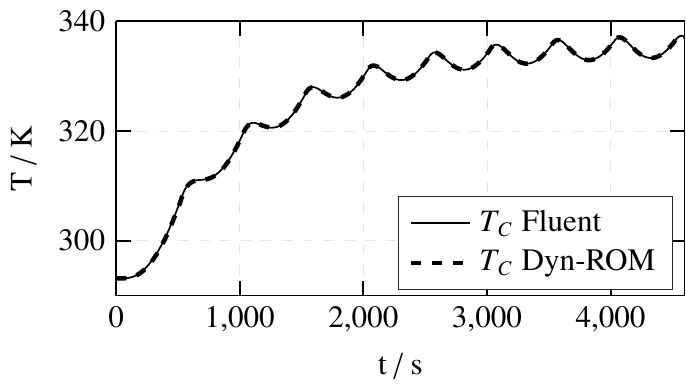}
\subcaption{Response to sawtooth signal input (trained).}
\label{fig:rom-dyn-examples-a}
\vspace*{0.5em}
\end{subfigure}
\begin{subfigure}[c]{0.48\textwidth}
\centering
\setlength\figureheightbig{0.2\textheight}
\setlength\figurewidthbig{0.98\textwidth}	
\input{tiks/garmod/ROM_SPThree473_outrom.tex} 
\subcaption{Response to three half-sine pulses (untrained).}
\label{fig:rom-dyn-examples-b}
\vspace*{0.5em}
\end{subfigure}
 \caption{Comparison of Dynamic ROM modeling and Fluent data.}
 \label{fig:rom-dyn-examples}
\end{figure*}

\subsection{Closed-loop control}
The performance of the presented Dynamic ROM in a closed-loop framework is evaluated. It will demonstrate that a controller design with moderate gain levels remains insensitive to the ROM model errors.
%
The Dynamic ROM is included in a simple PI control loop with conservative model gains $K_P=10$ and $K_I=0.01$ to avoid overshoots in this test. The setup is implemented in ANSYS Twin Builder 2019R2 (compare Fig.~\ref{fig:control-pid}) for the schematic signal flow. A limiter is added to avoid unphysical control signals, as the pan temperature is often limited to values of $500$~K maximum due to security reasons.
\begin{figure}[hbtp]
 \centering
\includegraphics[width=0.45\textwidth]{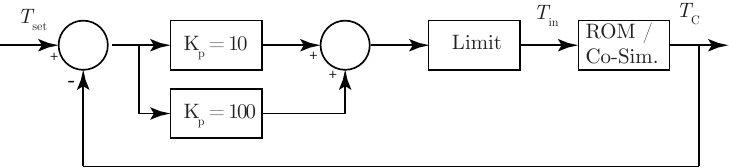}
 \caption{Concept of a control loop.}
 \label{fig:control-pid}
\end{figure}
Reference data is generated by including the full-order Fluent model in the control loop and controlling the model parameter $T_\text{in}$ via co-simulation.
The desired set-point $T_{set}= 330$~K is reached by both ROM and co-simulation, compare Fig.~\ref{fig:rom-pi-a}. The error between co-simulation and Dynamic ROM remains below 0.3~K. The small overshoot at $t=400$~s seems to be mitigated by the proportional controller part, as the prescribed input variable $T_\text{in}$ drops 2.5~K below the expected value. Considering the low error margin and the self-adjusting effect of the control loop, the Dynamic ROM is suitable for closed-loop control. 
%
\begin{figure*}[htbp]\small
\captionsetup[subfigure]{justification=centering}
\centering
\begin{subfigure}[l]{0.49\textwidth}
\includegraphics[width=0.99\textwidth]{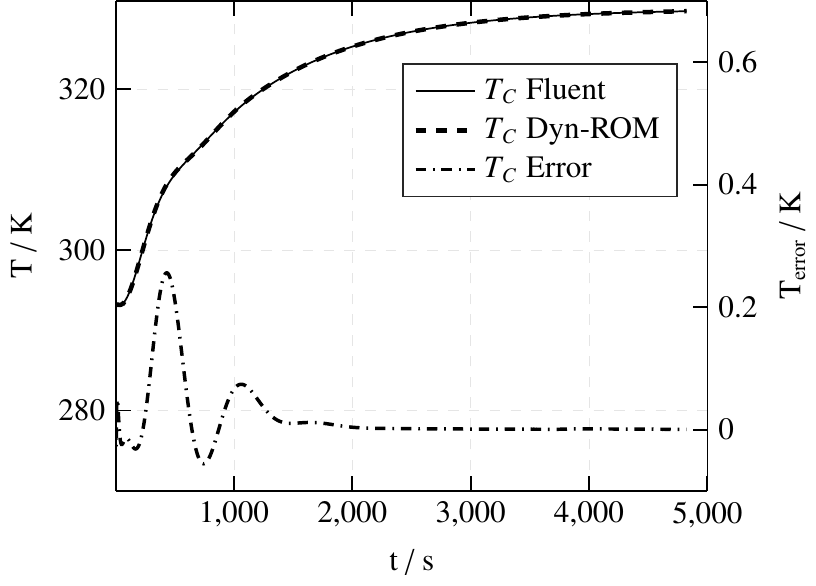}
\subcaption{Core temperature.}
\label{fig:rom-pi-a}
\end{subfigure}
\begin{subfigure}[c]{0.49\textwidth}
\centering
\includegraphics[width=0.99\textwidth]{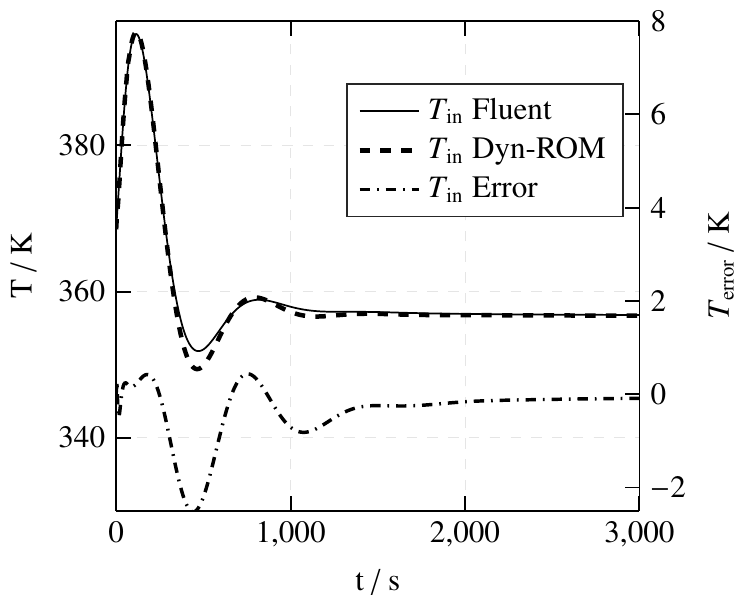}
\subcaption{Prescribed pan temperature by controller.}
\label{fig:rom-pi-b}
\end{subfigure}
 \caption{PI control of co-simulation and Dynamic ROM.}
 \label{fig:rom-PI}
\end{figure*}

\subsection{Reaction to a dynamic process environment}
The core concept of the Digital Twin methodology is real-time communication of simulation and process to positively influence the process under changing environmental conditions. As there is no direct communication interface to a cooking control software on the market yet, consider a simple academic scenario: In addition to the pan temperature, the ROM could be trained to adapt to changing convection conditions. One could assume that the central cooking control system knows the current fan speed of an industry-scale ventilation system. The additional forced convection would cool the chicken fillets and thus delay the desired cooking progress. The Digital Twin ROM can consequently predict the resulting implications on the core temperatures and adjust the pan temperature set-point accordingly. \begin{figure}[H]\small
\captionsetup[subfigure]{justification=centering}
\centering
\begin{subfigure}[c]{0.49\textwidth}
\centering
\setlength\figureheightbig{0.35\textheight}
\setlength\figurewidthbig{0.99\textwidth}	
\input{tiks/garmod/ROM_COSIM_Tc_vent2.tex}
\subcaption{Core temperature of chicken fillet with and without convection control.}
\label{fig:rom-vent-a}
\end{subfigure}
\begin{subfigure}[c]{0.49\textwidth}
\centering
\includegraphics[width=0.99\textwidth]{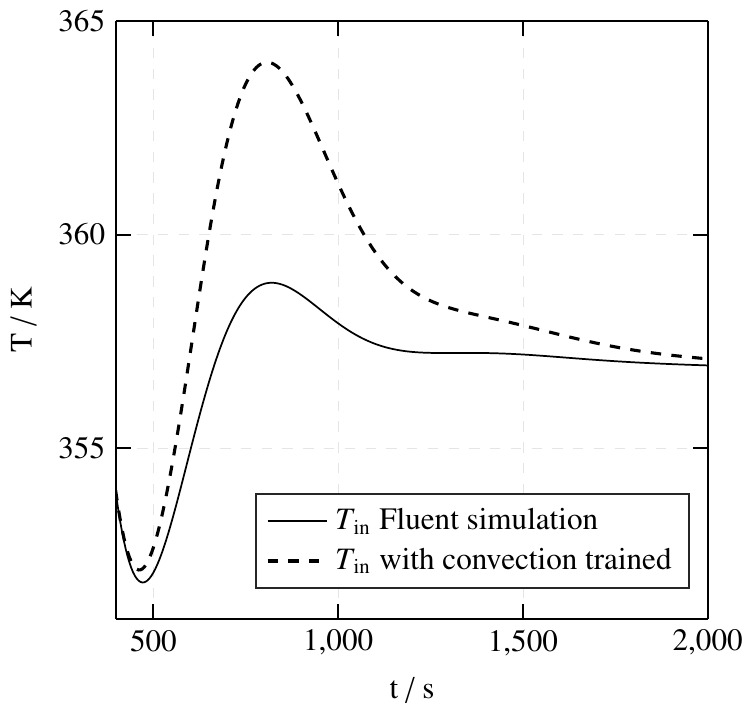}
\subcaption{Prescribed pan temperature by controller.}
\label{fig:rom-vent-b}
\end{subfigure}
 \caption{PI controller's reaction to increased convection.}
 \label{fig:rom-vent}
\end{figure}
Fig.~\ref{fig:rom-vent} depicts the reaction of the PI controller to an increased heat transfer to the ambient air. The onset of the forced convection is $t=400$~s, and its duration is 500~s. ROMs that are not trained to convection would follow a core temperature path with a delay of approximately two minutes, compare Fig.~\ref{fig:rom-vent-a}. On the contrary, a convective ROMs set-point is increased by up to $\Delta T = 6$~K, see Fig.~\ref{fig:rom-vent-b}. The control algorithm successfully mitigates the external influence and returns to the desired cooking path.

\section{Conclusion}
This proof-of-concept study demonstrates the successful use of Digital Twin methodology to control cooking processes.
Coupled transport equations for water concentration and temperature (denoted as C-T-model) are implemented to simulate the cooking of chicken meat. The results for the roasting process of chicken meat in a convection oven agree with experimental data ($E_\text{RMS,max}=1.87$~K). For pan-frying, the simulated temperatures match the average of nine frying experiments with an error band of $\pm 2$~K during the first ten minutes.
Non-intrusive reduced-order models are introduced to reduce the computational time to milliseconds. 
LTI models fail to predict cooling phases and heating effects beyond boiling temperature due to their non-linear nature. On the contrary, ANSYS Dynamic ROM can reproduce temperature responses to sophisticated excitation signals (such as sawtooth, rectangular or half-sine pulses) with significantly low error ($E_\text{max}=0.44$~K). Two examples demonstrate the usage of a Digital Twin to control the core temperature of the chicken. The PI closed-loop control system remains insensitive to errors induced by the Dynamic ROM evaluation.

This work marks the first steps of our research initiative to couple the heat transfer mechanisms of a cooking device to the culinary quality markers such as texture, gumminess, hardness and lightness~\cite{cm_feyissa_3D2013,cm_rabeler_kin2018} of food. 
In meat science, there is a long quest to describe the primary transport phenomena of $C$ and $T$ with the conservation of mass and energy. The correct water concentration and temperature representation are fundamental to successfully predicting culinary quality markers. However, sophisticated modelling of ``the other side of the food'' -- the cooking device and respective environment -- is under-represented in literature~\cite{cm_datta_porousmedia06a,cm_vandersman_moisture2007a,cm_vandersman_chickentunnel2013,cm_feyissa_phd2011,cm_feyissa_3D2013,cm_rabeler_mod2018}. 
Studies of heat transfer mechanisms in industry-scale cooking devices have revealed that already at temperatures of 170 to 250~$^\circ$C radiation effects cannot be ignored. They can account for up to 50 \% of the total heat flux~\cite{mscthesis}. Studies by Ibarra et al.~\cite{cm_ibarra_emissivity2000} found the chicken meat emissivity to be variable over time. Natural convection is a second, variable effect of heat loss. Due to the unsteady, turbulent flow over the food's topology, local changes in heat transfer coefficients are obvious (compare Fig~\ref{fig:vccacad_convection}). 
\begin{figure}[htbp]\small
\captionsetup[subfigure]{justification=centering}
\centering
\begin{subfigure}[c]{0.49\textwidth}
\centering
 \includegraphics[width=0.99\textwidth]{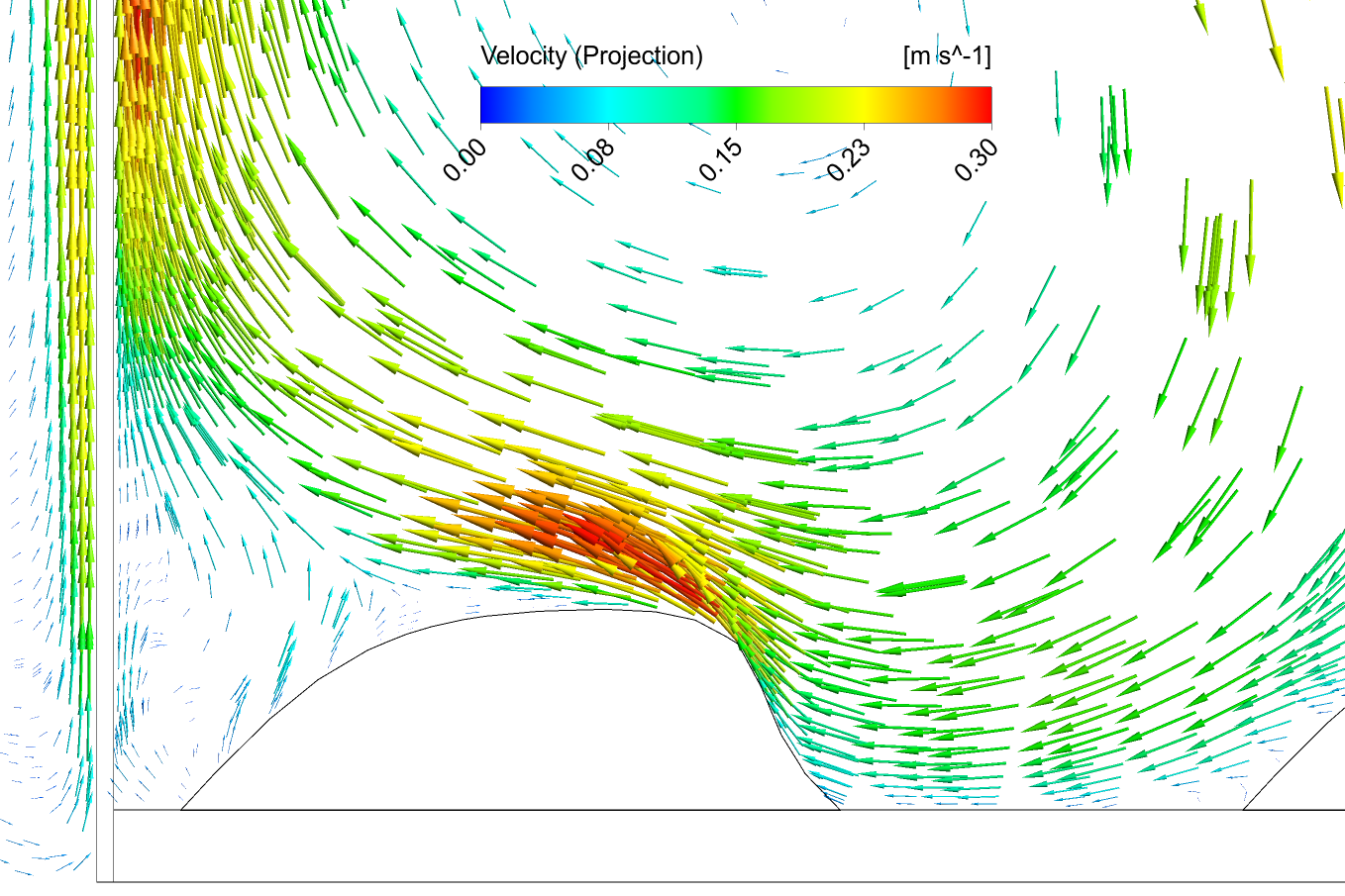}
\subcaption{Representative velocity of natural convection.}
\label{fig:vccacad_velovec}
\end{subfigure}
\begin{subfigure}[c]{0.49\textwidth}
\centering
 \includegraphics[width=0.99\textwidth]{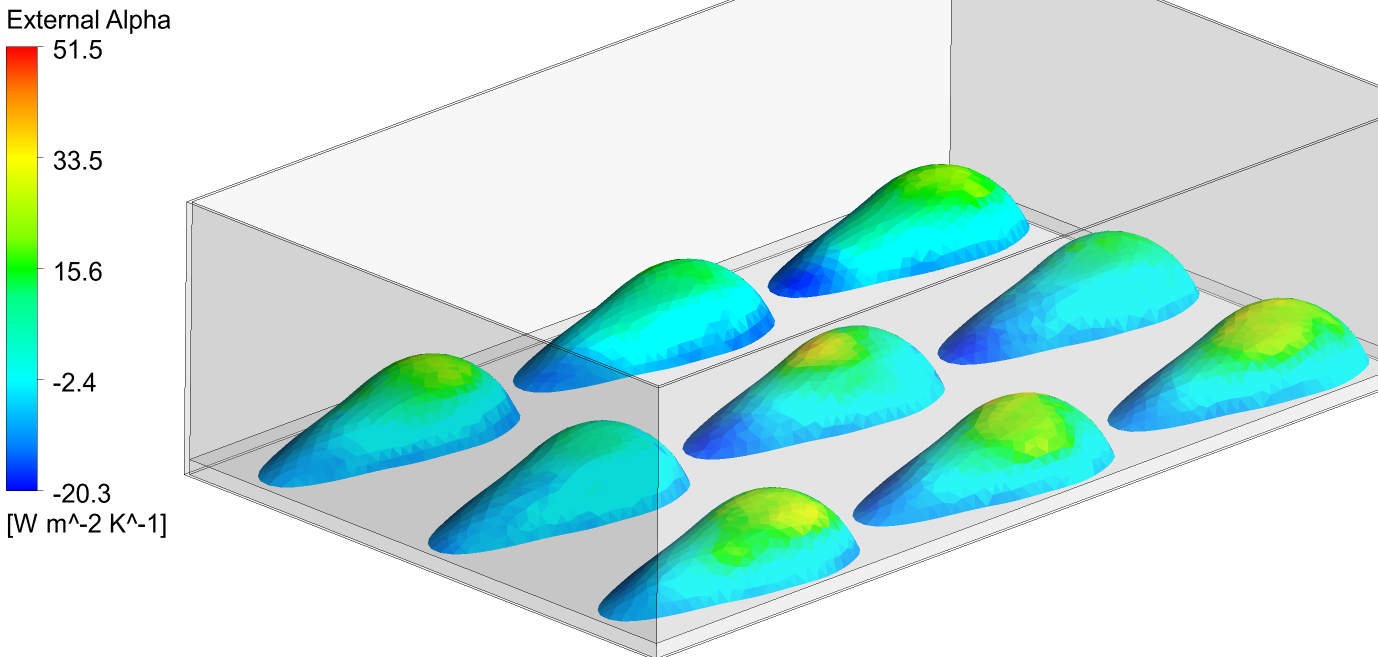}
\subcaption{External heat transfer coefficient at $t=600$~s.}
\label{fig:vccacad_alphaext}
\end{subfigure}
 \caption{Local variation of heat transfer due to natural convection.}
 \label{fig:vccacad_convection}
\end{figure}
However, most studies in the meat science community rely on averaged heat transfer coefficients. Herby, the chance to investigate heat transfer's unsteady and locally varying effects on the food's quality is missed. Future efforts will endeavor to close this gap on the path to intelligent and autonomous cooking devices.

 
 \section*{Acknowledgement}
 The work of Maximilian Kannapinn is supported by the ``Excellence Initiative'' of the German
Federal and State Governments and the Graduate School of Computational
Engineering at Technische Universität Darmstadt.

\appendix



\printbibliography



\end{document}